\documentclass[a4paper]{article}

\usepackage{INTERSPEECH_v2}
\usepackage{graphicx}
\usepackage{multirow}
\usepackage{hhline}
\usepackage{courier}
\usepackage{cite}

\title{Improving Source Separation via Multi-Speaker Representations}
\name{Jeroen Zegers, Hugo Van hamme}
\address{
  KU Leuven, Dept. ESAT, Belgium}
\email{jeroen.zegers@esat.kuleuven.be, hugo.vanhamme@esat.kuleuven.be}

\begin{document}

\maketitle
\begin{abstract}
Lately there have been novel developments in deep learning towards solving the cocktail party problem. Initial results are very promising and allow for more research in the domain. One technique that has not yet been explored in the neural network approach to this task is speaker adaptation. Intuitively, information on the speakers that we are trying to separate seems fundamentally important for the speaker separation task. However, retrieving this speaker information is challenging since the speaker identities are not known a priori and multiple speakers are simultaneously active. There is thus some sort of chicken and egg problem. To tackle this, source signals and i-vectors are estimated alternately. We show that blind multi-speaker adaptation improves the results of the network and that (in our case) the network is not capable of adequately retrieving this useful speaker information itself.
\end{abstract}
\noindent\textbf{Index Terms}: Source Separation, Single Channel, Blind Multi-Speaker Adaptation

\section{Introduction}
The cocktail party problem, where multiple sound sources, usually speakers, are simultaneously active, has been studied in the speech community for decades \cite{bregman1994auditory}. The problem is especially challenging when little assumptions are made. For a general solution we want to be speaker independent, text independent, using a single channel and so on. Recently, deep learning approaches have been used to address the cocktail party problem. To solve the problem, multiple speakers have to be segregated and thus an intra-class separation has to be made. Thus it is not possible to assign specific output nodes of the neural network to specific classes as is possible in for example speech-noise separation \cite{xu2014experimental,weninger2014single,wang2013towards}, male - female speech separation \cite{huang2014deep} or speaker dependent source separation \cite{huang2015joint}. If the model is to be trained speaker independently, a permutation problem is faced.

A way of permutation free learning was presented in \cite{Yu2016permutation}. They allowed all permutations during training and then only considered the one with the lowest loss. However at test time a hard decision had to be made on the permutation and some tracking was necessary to have consistent speaker assignments over a complete mixture. Other approaches to tackle the permutation problem consist of mapping each time-frequency bin (tf-bin) of a mixture spectrogram to an embedding space. Either an unsupervised clustering mechanism is then used to group tf-bins per speaker \cite{hershey2016deep,isik2016single} or the network learns attractors that draw points in the embedding space together and each speaker is represented by such an attractor \cite{chen2016deep}.

In speaker adaptation a system is adapted to better suit the characteristics of the target speaker. Speaker adaptation has been successfully applied is automatic speech recognition tasks using neural networks. There are several ways to incorporate speaker information in a network. One can apply a space transform to the input features depending on the speaker identity, such as maximum likelihood linear regression (MLLR) or feature-space MLLR (FMLLR) \cite{seide2011feature,yao2012adaptation}. The network can then be trained as usual. In model-based adaptation, the whole network or parts of the network are adapted to a specific speaker by retraining the model to adaptation data of that specific speaker \cite{swietojanski2014learning,liao2013speaker,yu2013kl,xue2014singular}. It is possible to perform so-called blind speaker adaptation by clustering speakers together via their i-vector representation \cite{zhang2011vector}. For each cluster a network is adapted. At test time an utterance is first assigned to a cluster and then decoded with the according network. The term blind is used, since the identity of the speaker is not known a priori, but still a network is used that is thought to be more adapted to the unknown speaker. Another way to adapt the network is to add speaker characterizing features, such as i-vectors \cite{dehak2011front} at the input \cite{saon2013speaker}. The advantage of such an i-vector is that it models speaker variability and can also be determined blindly, without any prior knowledge of the speaker \cite{gupta2014vector}.

In this paper an attempt is made to perform blind speaker adaptation for multi-speaker separation. There are two main challenges. The first challenge is that the network is to be adapted to more than a single person. Secondly, there is no direct way to extract an i-vector for all speakers, since they are speaking simultaneously. A multi-speaker representation is sought that can be added to the input of the network. The general idea is to first perform blind source separation, then extract i-vectors on the estimated sources and use these to adapt a second network, extract the i-vectors on the new estimates of the second network and so on. A similar idea was used by Zhang \textit{et al.} \cite{zhang2016pairwise} for noise suppression in presence of speech. They used speech enhancement to get a better estimate of the pitch, which they in turn fed in a subsequent network for better speech enhancement, which allowed for better pitch estimation and so on. If the i-vector extraction procedure is performed by a neural network, it is possible to do a final end-to-end training of the complete network. However, this will not be implemented in this paper. 

While i-vector extraction on signal estimates could allow for speaker verification in multi-speaker scenarios, the focus of this paper is on source separation quality. The rest of this paper is organized as follows. In section \ref{sec:SCSS} a brief overview of the baseline, a state-of-the-art multi-speaker source separation using neural networks, is given. In section \ref{sec:adap} a method is proposed to perform blind multi-speaker adaption for source separation. Experiments are presented in section \ref{sec:exp} and a conclusion is given in section \ref{sec:conc}.

\section{Single Channel Speaker Separation}
\label{sec:SCSS}
This section explains how a neural network, followed by a clustering algorithm can estimate a binary mask (BM) for each speaker, given a mixture. The framework of \cite{hershey2016deep} and partly of \cite{isik2016single} is used. A permutation problem arises when a neural network is asked to assign outputs nodes for each target speaker in the mixture. For example if a network is trained to output (A,B) when presented a mixture of speaker A and B and to output (A,C) when presented a mixture of speaker A and C, a problem arises when a mixture of speaker B and C is presented \cite{chen2016deep}. Therefor a network has to be trained with a permutation independent loss function. In \cite{hershey2016deep,isik2016single} this is done by mapping each time-frequency bin to an embedding space such that bins belonging to the same speaker are close together and those belonging to a different speaker are further apart. When assuming only one speaker is active per bin, an unsupervised clustering mechanism, like K-means, can be used to group bins and estimate a binary mask per cluster or target speaker.

Let $X_{i}, i \in \{1,\ldots,N\}$ be the short-time Fourier transform (STFT) of an audio mixture, with $i$ a time-frequency bin $(t,f)$, $N=TF$ and $T$ and $F$ the number of time frames and frequency bins, respectively. Project each tf-bin to a $D$-dimensional embedding space $V=f_{\theta}(X) \in \mathbb{R}^{N \times D}$ using a neural network. The used neural network is described in section \ref{sec:setup}. The embedding vector is normalized to unit length, so that $|v_i|^2=1$. Define an ($N\times C$)-dimensional target matrix $Y$, with $C$ the number of target speakers, so that $y_{i,c}=1$ if target speaker $c$ has the most energy in bin $i$ and $y_{i,c}=0$ if not. A permutation independent loss function (the columns in $Y$ can be interchanged without changing the loss function) can then be presented as
\begin{equation}
\mathcal{C}_Y(V)=||VV^T-YY^T||_F^2=\sum_{i,j}(\langle v_i,v_j\rangle-\langle y_i,y_j\rangle)^2.
\label{eq:loss}
\end{equation}
where $||A||_F^2$ is the squared Frobenius norm. Since $y_i$ is a one-hot vector,
\begin{equation}
\langle y_i,y_j\rangle=
\begin{cases}
      1, & \text{if}\ y_i=y_j \\
      0, & \text{otherwise}
    \end{cases}.
\end{equation}
The angle $\theta_{i,j}$ between the normalized vectors $v_i$ and $v_j$ is thus ideally 
\begin{equation}
\theta_{i,j}=
\begin{cases}
      0, & \text{if}\ y_i=y_j \\
      \pi/2, & \text{otherwise}
    \end{cases}.
\end{equation}
All embedding vectors $v_i$ are then clustered into $C$ clusters using K-means and hence a binary mask, $BM_c$, for each target speaker $c$ is created. The STFT of the estimated speech for speaker $c$ is then
\begin{equation}
\hat{S}_c=X*BM_c,
\end{equation}
where $*$ is an element wise multiplication.

\section{Blind multi-speaker adaptation}
\label{sec:adap}
\begin{figure}
\centering
\includegraphics[width=0.5\textwidth]{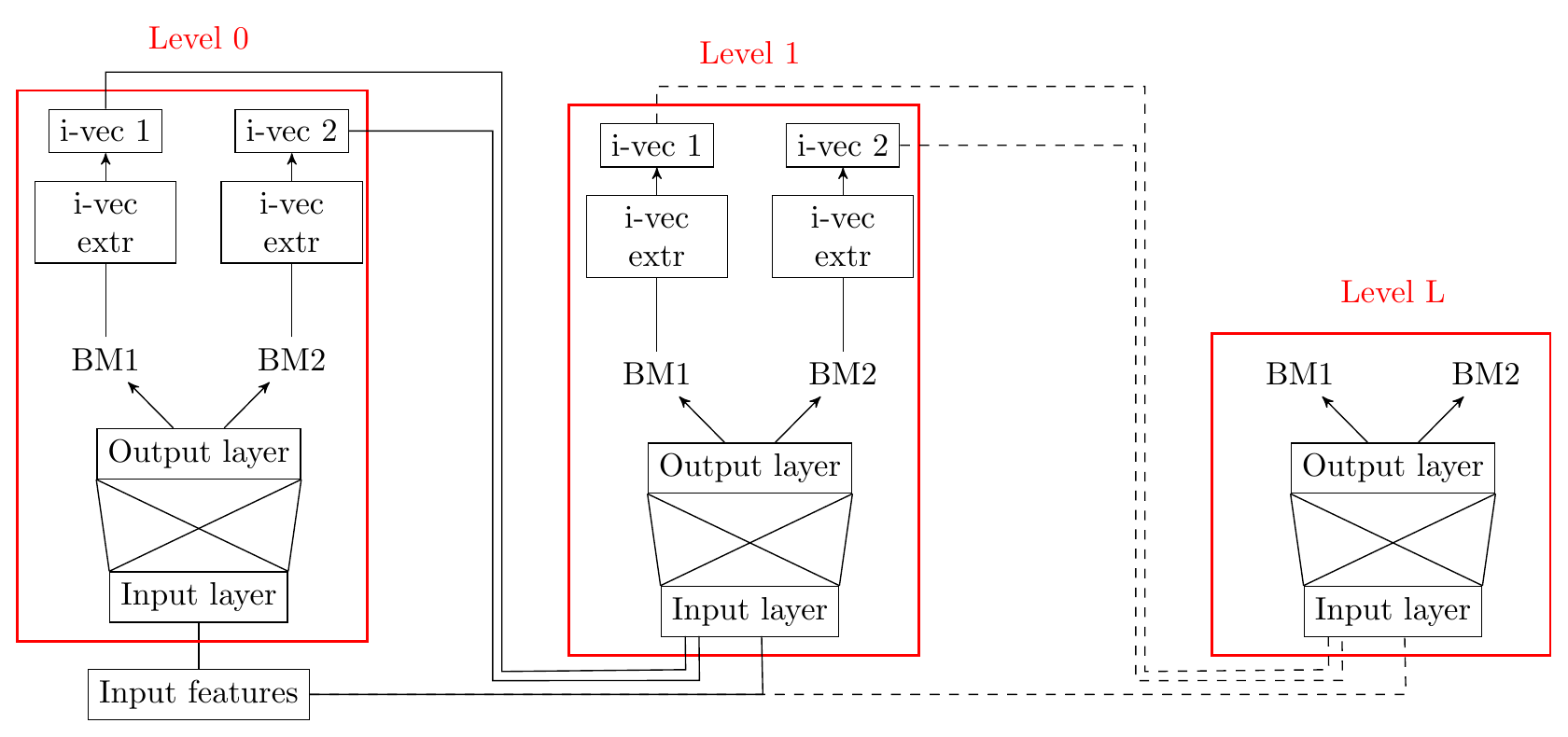}
\caption{Proposed iterative architecture \label{fig:arch}}
\end{figure}

To perform blind multi-speaker adaptation, first an estimate of the source signals is obtained as explained in section \ref{sec:SCSS}. Subsequently, an i-vector is extracted from each estimate. Optionally, the dimensionality of the i-vector can be reduced using Linear Discriminant Analysis (LDA). All (LDA) i-vectors, together with the mixture spectrogram, are then fed into another neural network. The last two steps can be repeated iteratively. The proposed iterative architecture is shown in figure \ref{fig:arch}. The network of level 0 is trained on the mixture spectrogram and is the state-of-the-art baseline for this paper. The network of level $l$ is trained on the mixture spectrogram and the (LDA) i-vectors of level $l-1$. Though one might expect improved performance with more levels $L$, no improvements were found beyond $L=1$ (see figure \ref{fig:res2}).

A Universal Background Model - Gaussian Mixture Model (UBM-GMM) is trained on development data. A supervector $M$ is derived for each utterance (e.g. from the estimates $\hat{S}_c$ of level $l-1$), using the UBM. $M$ is then represented by an i-vector $w$ and its projection based on the total variability space,
\begin{equation}
M=m+Tw,
\end{equation}
where $m$ is the UBM mean supervector, $w$ is the total variability factor or i-vector and $T$ is a low-rank matrix spanning a subspace with important variability in the mean supervector space and is trained on development data \cite{dehak2011front,glembek2011simplification}.


LDA can be used to further reduce the dimension of the i-vector. LDA is a rank reduction operation that tries to minimize intra-class variance and maximizes inter-class variance. Here, each class is represented by a single speaker. LDA tries to find the orthogonal directions $d$ that optimize the following ratio:
\begin{equation}
J(d)=\frac{d^tS_bd}{d^tS_wd}
\end{equation}
with $S_b$ the intra-speaker variance and $S_w$ the inter-speaker variance. The eigenvectors of the following eigenvalue equation are then found
\begin{equation}
S_bv=\lambda S_wv,
\end{equation}
where $\lambda$ is the diagonal matrix of eigenvalues. The eigenvectors with the highest eigenvalues are then stored in the projection matrix $A$ and the LDA i-vectors $w^*$ are then obtained as follows
\begin{equation}
w^*=A^Tw.
\end{equation}
Thus, a vector with inter-speaker discriminating dimensions is found.

These (LDA) i-vectors are obtained from the estimates of the previous neural network. The i-vectors are then stacked over all time frames of the mixture and fed into the new neural network, which is thus trained with extra speaker information.



\section{Experiments}
\label{sec:exp}
\subsection{Experimental set-up}
\label{sec:setup}
The proposed architecture was evaluated on two-speaker single-channel mixtures of the corpus introduced in \cite{hershey2016deep}, which contains 20,000 training mixtures ($\sim$ 30h), 5,000 validation mixtures (of which only 1,500 are used due to computing time restrictions) and 3,000 test mixtures. The mixtures were artificially mixed using utterances of the Wall Street Journal 0 (WSJ0) at various signal-to-noise ratios, randomly chosen between 0dB and 10 dB and sampled at 8 kHz. The training and validation set were constructed using the \texttt{si\_tr\_s} set and the evaluation set consists of 16 held-out speakers of the \texttt{si\_dt\_05} set and the \texttt{si\_et\_05} set. Since some form of blind speaker adaptation is performed, using the same speakers in the training and validation set might lead to some over fitting, but since evaluation is done on held out speakers, evaluation results are expected to only get better should the validation set also contain held out speakers. The magnitude of the STFT with a 64ms window length and 16ms hop size was used at the network's input, using mean and variance normalization, obtained over the whole training set. 

The \texttt{si\_tr\_s} set of Wall Street Journal 1 (WSJ1) was used as development data to train the UBM, $T$ and $A$. 13-dimensional Mel-Frequency Cepstral Coefficients (MFCC's) are used as features and a VAD was used to leave out the silence frames. The UBM has 256 mixtures and $w$ is 400-dimensional, unless mentioned otherwise. The dimensionality of $A$ is a tunable parameter in the experiment section. The MATLAB MSR Identity Toolbox v1.0 \cite{sadjadi2013msr} was used to determine the UBM, $T$ and $A$ and to obtain the (LDA) i-vectors.

The neural network has 2 fully connected BLSTM layers with 600 hidden units each with a tanh activation \cite{hochreiter1997long}. The embedding dimension $D$ was set to 20 so that the linear output layer had $FD=256*20=5120$ units. The input dimension was equal to $F=256$. When (LDA) i-vectors were used the input dimension increased to $(F+C*ivec\_dim)$. The Adam learning algorithm was used with initial learning rate $10^{-3}$, $\beta_1=0.9$, $\beta_2=0.999$ and $\epsilon=10^{-8}$ \cite{kingma2014adam}. Unlike in \cite{isik2016single}, dropout on the feedforward weights did not improve results, so it was not used in the experiments. The batch size was taken at 128 and every 10 batches the validation loss of equation \ref{eq:loss} was calculated. If the validation loss increased, the previous validated model was restored and the learning rate was halved. Early stopping was applied when validation loss increased 3 consecutive times. Zero mean Gaussian noise with standard deviation 0.6 was applied to the training inputs (including the i-vectors for the speaker adapted networks).
Tf-bins with magnitude -40 dB, compared to the maximum of the utterance, were omitted in the loss function of equation \ref{eq:loss} to prevent the network from learning on empty or low-energy bins. In the experiments below, for every architecture, six independent runs were used and the network with the lowest validation loss was kept. This was to cope with some variability due to the applied input noise and the random initialization of the networks (the latter did not apply if the network was initialized with another network). The networks were trained using curriculum learning \cite{bengio2009curriculum}, i.e. the networks were presented an easier task before tackling the main task. Here, the network was first trained on 100-frame non-overlapping segments of the mixtures. This network was then used to initialize for training over the full mixture.  All networks were trained using TensorFlow \cite{abadi2016tensorflow}.

The K-means clustering was done with the built-in MATLAB function using the cosine distance and 10 random initializations, choosing the version with lowest total sum of distances. Again, tf-bins with magnitude -40 dB, compared to the maximum of the utterance were omitted.

\subsection{Results}
\begin{figure}
\centering
\includegraphics[width=0.5\textwidth]{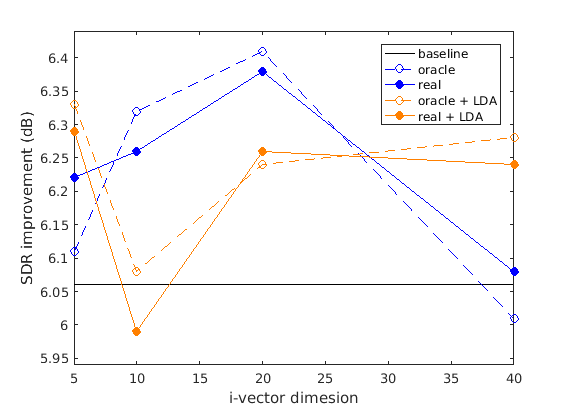}
\caption{\label{fig:res1}{SDR improvements for oracle and realistic experiments (in dB)}}
\end{figure}

Performance was measured in signal-to-distortion ratio (SDR) improvements on the evaluation set, using the \texttt{bss\_eval} toolbox \cite{vincent2006performance}. The average SDR of the mixture was 0.15dB. The baseline system did not use any speaker adaptation and was based on section \ref{sec:SCSS}. Results are comparable to \cite{hershey2016deep}. Results for the multi-speaker adapted networks are shown in figure \ref{fig:res1}. For the oracle experiments, (LDA) i-vectors were obtained from the original single speaker utterances, which are normally not available at test time. Results are shown for i-vectors with different dimensions. For the LDA i-vectors, the original i-vectors were 400-dimensional and LDA was used to reduce the dimension. Improvements up to 0.35 dB were found by adding this simple multi-speaker representations. These oracle networks were then used to initialize the realistic networks. For these realistic networks, i-vectors are used which are
obtained from the signal estimates produced by the baseline network, both for training and testing. The realistic results are similar to the oracle results, up to 0.32 dB increase compared to the baseline. This seems to indicate that the network learns to cope with the fact that for the realistic experiments, i-vectors are not obtained from clean single-speaker utterances but rather from estimated source signals. Also, it might not be ideal to try to represent both speakers separately and concatenating both representations. Possibly, a true multi-speaker representation, where a single representation is used for the combination of the two speakers, would allow for further improvements. 


Results are not too much dependent on the dimensionality of the i-vector. This can be explained by the fact that the first dimensions contain the most variation, both for the i-vectors as for the LDA i-vectors. There is also no big difference between i-vectors and LDA i-vectors. 
This may indicate that the first few components of the i-vectors reflect mostly inter-speaker variation rather than channel effects and the like. Also, a representation of intra-speaker variability can be useful for source separation while it is expected that LDA would remove this information.

\begin{figure}
\centering
\includegraphics[width=0.5\textwidth]{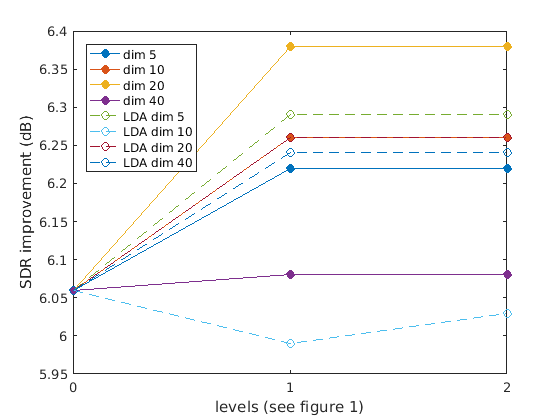}
\caption{\label{fig:res2}{SDR improvements when using multiple levels for (LDA) i-vector extraction (in dB)}}
\end{figure}


In figure \ref{fig:res2} it is shown how results further improve when using the outputs of the first level to again estimate new (LDA) i-vectors and use these at the input of the second level network, which was initialized using the level 1 network. Since the separation quality of the first level is better then the baseline, i-vectors are expected to be a better representation of a speaker. However, the increase from 1 to 2 levels is very minimal. There is only a small difference between the oracle and realistic experiments in figure \ref{fig:res1} and thus \textit{cleaner} i-vectors do not make a big difference. Furthermore, the increase of separation quality of the first level compared to the baseline is only limited and thus no big difference in i-vector extraction quality is expected.


\subsection{Speaker identification and representation analysis}
The focus of this paper was on source separation quality. However, because i-vectors were obtained in the proposed architecture, it is checked how they would cope in speaker identification experiments. Note, however, that the i-vectors were chosen as extra input features for the neural net and they are not optimized towards speaker identification accuracy. 
In the test set there were 9 speakers, for which at least 5 of their utterances from the WSJ0 database were not used to create test mixtures. For each of these 9 speakers, (LDA) i-vectors were obtained for these left-over utterances and speaker models were created by averaging i-vectors per speaker. An i-vector was also obtained for each source signal that was estimated by a network. It was then classified to a test speaker i-vector model using the cosine distance. Only source signal estimates belonging to one of these 9 speakers were considered. Speaker identification accuracy is shown in table \ref{tab:SR1}. Notice however, that the speaker i-vector models were obtained on clean utterances, while test i-vectors were obtained from estimated source signals, which explains the low identification accuracy. 
To clarify, the baseline results in table \ref{tab:SR1} refer to i-vectors extracted by a level 0 network. The oracle results refer to the i-vectors extracted by a level 1 network, when oracle i-vectors were used at the input. The real results refer to the i-vectors extracted by a level 1 network, when i-vectors extracted by the level 0 network were used at the input.

\begin{table}
\centering
  \begin{tabular}{c | c c c | c c c}
  &  \multicolumn{3}{c}{i-vector} &  \multicolumn{3}{|c}{LDA i-vector} \\
  \hhline{~------}
     & baseline & oracle & real & baseline & oracle & real \\
   \hline
  dim=5 & 42.8 &  44.9 & 43.5 & 24.4 & 25.1 & 24.3\\ 
  dim=10 & 30.2 & 32.6 & 30.8 & 29.6  & 26.9 & 30.3 \\ 
  dim=20 & 22.3 & 22.5 & 21.4& 19.7  & 20.3 &  19.3 \\ 
  dim=40 & 39.4 & 38.7 & 37.7 &  24.6 & 26.0 & 24.8  \\ 
  \end{tabular}
    \caption{\label{tab:SR1}{Identification accuracy (in \%) of speakers in multi-speaker mixtures}}
  \end{table}

In most cases, i-vectors obtained from the \emph{real} networks achieve better speaker identification accuracy then those deduced from the baseline network. This was expected since SDR was also higher and thus the estimates are closer to the original signals.

Finally, a small experiment was done to analyze the quality of representation of the i-vector. The i-vectors obtained from the output of the baseline are used at the input for the \emph{real} level 1 network experiments. It is expected that a \emph{good} i-vector representation of a speaker, will allow for better source separation. If the representation is \emph{insufficient}, it is expected that the source separation using the oracle i-vector representations will be better. The i-vector representation, obtained from the baseline network, is said to be \emph{sufficient} when it is classified to the correct test speaker in the speaker identification experiment. In table \ref{tab:SR2} the average level 1 SDR improvement of signal estimates where the speaker was correctly identified is shown, as well as the average SDR improvement of signal estimates for falsely identified speakers. The table also shows the average SDR increase when oracle representations were used instead of baseline i-vectors. Due to space constraint only experiments for LDA i-vectors are shown.

\begin{table}
\centering
  \begin{tabular}{c | c c | c c}
     & corr ID &  oracle incr & false ID & oracle incr\\
   \hline
  dim=5 & 6.04 &  +0.04 & 6.12 & +0.05 \\ 
  dim=10 & 5.76 & +0.12 &  5.85  & +0.08 \\ 
  dim=20 & 6.02 & -0.07 & 6.06 & +0.01 \\ 
  dim=40 & 6.32 & +0.03 & 5.98 & +0.12  \\ 
  \end{tabular}
    \caption{\label{tab:SR2}{Average SDR improvement (in dB), depending on correct LDA i-vector representation, and the difference with the oracle experiments.}}
  \end{table}

It is noticed that usually the SDR increase for the oracle experiments is bigger for utterances that had an \emph{insufficient} i-vector representation. The above assumption, that a better speaker representation leads to better separation quality, thus holds to some extent. However, because SDR of the \emph{oracle} experiments in figure \ref{fig:res1} were not much higher than the \emph{real} experiments, differences are very small and might not be statistically significant.


\section{Conclusions}
\label{sec:conc}
In this paper it was shown that the presented neural network baseline is not adequately capable of extracting speaker information in multi-speaker mixtures for a source separation task. It was shown that explicitly extracting this speaker information (using a baseline network) and adding this information to the input of the network, improves results. Improvements of about 0.3 dB were found. Further research should point out whether this gain changes when deeper or different architectures are used. 

An initial attempt was made to make a multi-speaker representation. In this paper, two single-speaker representations were concatenated to make such a multi-speaker representation. Better results are expected when a single representation is directly used for the combination of two speakers. Especially if the representation focuses on the difference between the two speakers. This seems very relevant when performing speaker segregation.

\section{Acknowledgements}
This work was funded by the SB PhD grant of the Research Foundation Flanders
(FWO) with project number 1S66217N and the KULeuven research grant
GOA/14/005 (CAMETRON).

\bibliographystyle{IEEEtran}
\bibliography{mybib}

\end{document}